\documentclass[12pt]{article}
\usepackage{graphicx} 
\usepackage{amsmath,amsmath, amssymb}
\usepackage{natbib}
\usepackage{url} 

\newcommand{\blind}{0}

\usepackage{color,tikz,xcolor}
\usepackage{multirow}
\usepackage{pdflscape}
\newdimen\nodeDist
\usepackage[colorlinks,citecolor=blue,urlcolor=blue,filecolor=blue,backref=page]{hyperref}
\usepackage{caption, subcaption}
\usepackage{algorithm, algorithmicx,algpseudocode}
\usepackage{booktabs}
\usepackage{bbm}

\usepackage{todonotes}
\newcounter{todocounter}

\newcommand{\x}{\mathrm{x}}

\newcommand{\indep}{\perp\!\!\!\perp}

\def\spacingset#1{\renewcommand{\baselinestretch}%
{#1}\small\normalsize} \spacingset{1}

\begin{document}

\if0\blind
{
  \title{\bf LongBet: Heterogeneous Treatment Effect Estimation in Panel Data}
  \author{Meijia Wang\hspace{.2cm}\\
    Google\hspace{.2cm}\\
    Ignacio Martinez\hspace{.2cm}\\
    Google\hspace{.2cm}\\
    P. Richard Hahn\hspace{.2cm}\\
    Arizona State University}
    \date{}
  \maketitle
} \fi

\if1\blind
{
  \bigskip
  \bigskip
  \bigskip
  \begin{center}
    {\LARGE\bf LongBet: Heterogeneous Treatment Effect Estimation in Panel Data}
\end{center}
  \medskip
} \fi

\bigskip
\begin{abstract}
This paper introduces a novel approach for estimating heterogeneous treatment effects of binary treatment in panel data, particularly focusing on short panel data with large cross-sectional data and observed confoundings. In contrast to traditional literature in difference-in-differences method that often relies on the parallel trend assumption, our proposed model does not necessitate such an assumption. Instead, it leverages observed confoundings to impute potential outcomes and identify treatment effects. The method presented is a Bayesian semi-parametric approach based on the Bayesian causal forest model, which is extended here to suit panel data settings. The approach offers the advantage of the Bayesian approach to provides uncertainty quantification on the estimates. Simulation studies demonstrate its performance with and without the presence of parallel trend. Additionally, our proposed model enables the estimation of conditional average treatment effects, a capability that is rarely available in panel data settings. 
\end{abstract}

\noindent%
\vfill

\newpage

\section{Introduction}

Causal inference problems in panel data involve understanding and quantifying the causal relationships between the response and the treatment over time. It is widely studied in policy evaluation and intervention analysis. This allows policymakers to study the effect of a policy on target variables at different time points, considering the temporal ordering of events. Panel data can provide valuable insights into dynamic processes of the causal effect and heterogeneity on different units.

Here we focus on the study of heterogeneous time-varying treatment effects in panel data. Time-varying treatment effects refer to the situation where the effect of a treatment or intervention may change over time, potentially exhibiting different patterns or magnitudes at different points in time. The key to the problem is to effectively estimate the potential outcomes for the treated groups and identify the treatment effect given observed treated outcomes. 

Several methods have been developed in the Economics literature to study the average treatment effects over time with different settings. Fixed effects methods \citep{CALLAWAY2023184, imbens2021controlling} model the response variable by fixed effects for individuals and factor loadings over time and use it to impute the unobserved potential outcomes for the treated. Relying on the parallel trend assumption, difference-in-differences design \citep{CALLAWAY2021200, borusyak2023revisiting, Abadie2005SemiDID} imputes the potential outcomes for the treated parallel to observed control units based on the pre-treatment differences. \citep{Abadie2021UsingSC} provides an overview of the synthetic control method, which predicts the potential outcome of each treated unit with an ensemble of observed control units. Recent works \citep{Acemoglu2013TheVO, kreif2016sc, abadie2021penalizedsc, dube2015poolsc} have applied the synthetic control approach to a large number of units. Interrupted time series analysis \citep{Bernal2016InterruptedTS, LopezBernal2018AMF, Schaffer2021InterruptedTS} focuses on studying the effect of population-level interventions by analyzing the time series data before and after the intervention. It involves modeling the trend and autocorrelation in the outcome variable while considering the timing of the intervention. Other methods include the matrix completion approach \citep{Athey_2021}.

While most literature focuses on studying average treatment effects over time for a treatment adopted at a particular time, recent works have been developed to accommodate staggered adaption and estimate heterogeneous treatment effects. \citep{borusyak2023revisiting} accommodate the heterogeneity in treatment effect by leveraging the fixed effect model. \citep{CALLAWAY2021200} relaxes the parallel trend assumption to hold only after conditioning on observed covariates. \citep{jeffrey2021twmr} proposed a two-way Mundlak (TWM) regression that allows for heterogeneity in treatment trends.  \citep{xu_2017} proposes a generalized synthetic control method that relaxes the parallel trend assumption and allows for multiple treatment units and varying treatment by combining synthetic control with a linear fixed effects model but requires observed covariates to be time-varying. Xu also proposed other methods (to be cited) that require time-varying covariates. \citep{Athey_2021, tanaka2020bayesian} propose using the matrix completion method to impute potential outcomes as missing values and in turn estimate the treatment effect.

Traditional methods in causal inference often rely on the parallel trend assumption, assuming that the untreated and treated groups follow parallel trends over time. However, in various real-world scenarios, this assumption may not hold, leading to limitations in the applicability of traditional methods. For instance, in e-commerce, different verticals may exhibit seasonality, and user-level data can vary significantly, making it challenging to satisfy the parallel trend assumption. Similarly, in patient analysis, individual responses to treatments may diverge, rendering the parallel trend assumption inappropriate. Nevertheless, these datasets often provide valuable observed covariates that can effectively characterize individual behaviors and affect treatment responses. As a result, we propose a new method that can leverage these observed covariates to impute potential outcomes and estimate treatment effects, offering a powerful and flexible approach that can address these complex scenarios without rigorously relying on the parallel trend assumption.

\citep{hill2011bayesian} first advocates using Bayesian Additive Regression Trees (BART) in causal inference. BART's flexibility and robustness to hyperparameters make it an appealing choice for handling complex treatment effect heterogeneity. Building on this foundation, Bayesian Causal Forest (BCF) \citep{richard2020bcf} models and regularizes the heterogeneous treatment effects separately from the prognostic effect of control variables using two BART forests in cross-sectional settings. Additionally, \citep{Krantsevich2022StochasticTE} introduced the XBCF model which inherits the modeling of BCF with significantly accelerated algorithm. Inspired by the BART literature, we propose a novel approach for treatment effect estimation in panel data, thereby offering a unique and distinctive approach that sets our work apart from existing literature in the field.

Here we consider a scenario where we observe panel data of the response to a binary treatment. We observe pre-treatment confoundings that affect the treatment adoption and the treatment effect, which are observed. We consider a "staggered rollout" design where the units can be treated at different times but once a unit is treated it remains treated afterward. The goal is to estimate the heterogeneous treatment effect conditioning on observed confoundings. As a by-product, we can also provide an estimation of the average treatment effect and/or average treatment effect on the treated by averaging over the conditional treatment effect. 

Our proposed method offers several distinct advantages over traditional approaches in estimating time-varying treatment effects. Unlike methods that rely on the parallel trend assumption, our model does not impose such restrictive assumptions, allowing for greater applicability in scenarios with diverse and complex data patterns. Moreover, our model is the first to harness the power of Bayesian Additive Regression Trees (BART) for capturing time-varying treatment effects, providing a novel and versatile framework. As a semi-parametric model, it offers flexibility in capturing treatment effect heterogeneity, accommodating variations in data across different time periods. The model also readily handles staggered adoption of treatments, making it suitable for a wide range of real-world applications. Leveraging a Bayesian approach, our method provides credible intervals for conditional treatment effects, enhancing the interpretability and reliability of the results. Collectively, these features make our approach a powerful and innovative tool for causal inference in the dynamic context of time-varying treatment effects.

\section{Background}
\subsection{Bayesian additive regression trees (BART)} 
Bayesian Additive Regression Trees (BART), proposed by Chipman et al. \citep{chipman2010bart}, is a tree-based regression model known for its remarkable capacity to model complex response surfaces, particularly in situations characterized by large and intricate covariate structures. This model operates through an ensemble of decision trees, providing both flexibility and interpretability. In addition to its predictive power, BART is equipped with the capability to offer Bayesian credible intervals, rendering it an invaluable tool for conducting inference. 

The BART model models the regression problem as the following:
\begin{equation}
y_i = \sum_{l=1}^{L} g(x_i; T_l, \mu_l) + \epsilon_i,
\end{equation}
which seeks to approximate the relationship between covariates $\mathbf{x}$ and the response variable $y$ through a composition of regression trees. Here, $T_l$ denotes a tree structure consisting of a set of partitioning rules that divide the covariate space into distinct subspaces, represented as $\{\mathcal{A}_{1}, \cdots, \mathcal{A}_{B}\}$. Correspondingly, $\mu_l = (\mu_{l1}, \cdots, \mu_{lB})$ constitutes a vector of leaf parameters, each associated with the leaf nodes within tree $T_l$. 

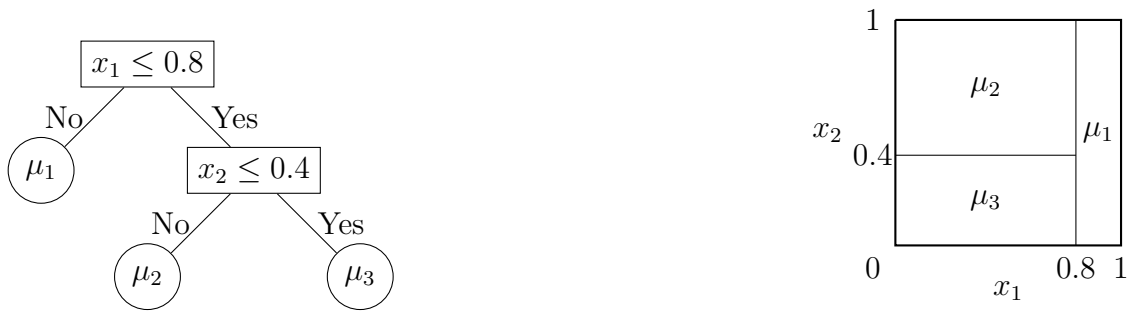
\begin{figure}[!ht]
	\begin{subfigure}{171pt}
		\begin{center}
			\begin{tikzpicture}[
					scale=0.8,
					node/.style={%
							draw,
							rectangle,
						},
					node2/.style={%
							draw,
							circle,
						},
				]
				\node [node] (A) {$x_1\leq0.8$};
				\path (A) ++(-135:\nodeDist) node [node2] (B) {$\mu_{1}$};
				\path (A) ++(-45:\nodeDist) node [node] (C) {$x_2\leq0.4$};
				\path (C) ++(-135:\nodeDist) node [node2] (D) {$\mu_{2}$};
				\path (C) ++(-45:\nodeDist) node [node2] (E) {$\mu_{3}$};

				\draw (A) -- (B) node [left,pos=0.5] {No}(A);
				\draw (A) -- (C) node [right,pos=0.5] {Yes}(A);
				\draw (C) -- (D) node [left,pos=0.5] {No}(A);
				\draw (C) -- (E) node [right,pos=0.5] {Yes}(A);
			\end{tikzpicture}
		\end{center}

	\end{subfigure}
	\hfill
	\begin{subfigure}{171pt}

		\begin{center}
			\begin{tikzpicture}[scale=3]
				\draw [thick, -] (0,1) -- (0,0) -- (1,0) -- (1,1)--(0,1);
				\draw [thin, -] (0.8, 1) -- (0.8, 0);
				\draw [thin, -] (0.0, 0.4) -- (0.8, 0.4);
				\node at (-0.1,0.4) {0.4};
				\node at (0.8,-0.1) {0.8};
				\node at (-0.1, -0.1) {0};
				\node at (1, -0.1) {1};
				\node at (-0.1,1) {1};
				\node at (0.5,-0.2) {$x_1$};
				\node at (-0.3,0.5) {$x_2$};
				\node at (0.9,0.5) {$\mu_{1}$};
				\node at (0.4,0.7) {$\mu_{2}$};
				\node at (0.4,0.2) {$\mu_{3}$};
			\end{tikzpicture}
		\end{center}
	\end{subfigure}
	\caption{A regression tree partitions a two-dimensional covariate space (right) with two internal decision nodes (left). Each element of the partition corresponds to a terminal node in the tree with leaf parameter $\mu_l$.}\label{fig:tree_example}
\end{figure}

The binary regression tree shown in Figure \ref{fig:tree_example} serves as an illustrative example. This tree encompasses a collection of internal decision nodes as well as terminal nodes, often referred to as leaf nodes. These leaf nodes effectively partition the covariate space into a total of $B$ segments. The function of the tree, denoted as $g(\mathbf{x}, T_l, \mu_l)$, adopts a step-like structure. This function allocates a point $\mathbf{x}$ to a leaf node and assign the corresponding leaf parameter according to the split rules in the tree $T_l$.

At the core of BART lies the formulation of its regularized tree prior and leaf parameter prior, which characterizes its regularized nature while still being flexible to model complex response surfaces. The tree prior put a strong regularization on tree depth, the likelihood to split a node decreases exponentially as the tree grows deeper. The leaf parameter prior assumes a normal distribution, indicating that the model assumes that the responses within each leaf node follow a Gaussian distribution. This encourages smoother estimates within individual leaves, which, in turn, contributes to the overall smoothness of the model's predictions. 

BART employs a random walk Metropolis-Hastings (MH-MCMC) algorithm to generate trees from the posterior distribution. During each iteration, the algorithm proposes a single growth or pruning procedure for each tree. The decision to accept or reject these proposals is determined through a Metropolis-Hastings (MH) approach, where the posterior probability of the proposed split is evaluated. Furthermore, the iterative nature of BART involves fitting each tree on the residuals of the ensemble formed by the other trees. This approach ensures that subsequent trees focus on capturing the remaining patterns in the data that have not been explained by the existing ensemble. 

However, the algorithm's efficiency diminishes when handling larger datasets or covariate spaces. In contrast, a more recent approach known as XBART \citep{he2019xbart, he2023stochastic} draws inspiration from BART but adopts a different strategy. Instead of making incremental modifications to pre-existing trees, XBART generates entirely new trees with each iteration. At each node, the algorithm explores a comparable range of cutpoints as BART and samples split criteria proportionate to the marginal likelihood. This methodology facilitates the more efficient discovery of well-fitting deep trees compared to BART MH-MCMC and performs better when handling larger datasets. Our proposed method aligns with both XBART and BART frameworks, although we will primarily demonstrate its efficacy using XBART for computational efficiency.

\subsection{Bayesian Causal Forest (BCF)} 
Constructed upon the foundation of BART model, the Bayesian Causal Forest (BCF) model stands as a robust Bayesian non-parametric technique designed for estimating heterogeneous treatment effects in cross-sectional data. Its effectiveness has been demonstrated through applications on simulated and real-world datasets \citep{richard2020bcf}, showcasing superiority over popular alternatives like BART, causal random forests, and propensity score matching methods.

BCF identifies treatment effect based on the potential outcome framework \citep{Imbens2015CausalIF} and relies on the assumption of the Stable Unit Treatment Value Assumption (SUTVA) and the Strong Ignorability Assumption. SUTVA stipulates that a unit's potential outcome depends solely on its own treatment status, unaffected by the treatments of others. Meanwhile, the Strong Ignorability Assumption asserts that given observed covariates, treatment assignment is independent of potential outcomes.

The BCF model formulates the response variable $Y$ as the sum of a prognostic function and a treatment function given binary response $z_i$, represented as:
\begin{equation} \label{eq:bcf}
    \mathbb{E}(Y_i | \x_i, Z_i = z_i ) = \mu(\x_i, \hat{\pi}(\x_i)) + \tau(\x_i) z_i.
\end{equation}
Here, $\mu$ captures the prognostic impact, and $\tau$ represents the treatment effect. Both of these functions are estimated through BART forests, which empowers the model to effectively approximate complex response surfaces. Furthermore, \citep{richard2020bcf} suggested incorporating the estimated propensity score $\hat{\tau}(\x_i)$ in the prognostic effect estimation, a strategy that has showcased enhanced accuracy.

XBCF \citep{Krantsevich2022StochasticTE} is a recent innovation rooted in the Bayesian Causal Forest \citep{richard2020bcf}. This model expresses observable outcomes as:
\begin{equation} \label{eq:xbcf}
\begin{aligned}
   y_i &= a \mu(\x_i, \hat{\pi}_i)) + b_{z_i} \tau(\x_i) + \epsilon_i, \quad  \epsilon_i \sim N(0, \sigma^2_{z_i}),\\ 
   &a\sim N(0,1), \quad b_0, b_1\sim N(0, 1/2)
\end{aligned}
\end{equation}
It extends the foundational Bayesian Causal Forest (BCF) technique with the aim of boosting its performance in heterogeneous treatment effect estimation and computational efficiency. While BCF employs the MH-MCMC algorithm from BART to estimate both prognostic and treatment functions, XBCF leverages the structure of XBART to enhance the efficiency of fitting the BCF model. The model's effectiveness is illustrated through comparisons with related approaches in both simulation studies and empirical analyses. 

The two models motivated us to extend their applicability to causal panel data, thereby offering a unique and distinctive approach that sets our work apart from existing literature in the field.

\section{Methodology}
In this section, we will begin by formalizing the problem, focusing on time-varying treatment effect of a binary treatment in panel data. The binary treatment can be applied to the treatment group at a specific time or be implemented in a staggered roll-out design. Then, in order to identify treatment effect, we will employt the dynamic potential outcomes framekwork and extend the strong ignorability assumption to panel data. Finally, we will introduce our proposed model LongBet.

Let $Y_{it}$ represent the continuous response of a unit $i$ at time $t$, $Z_{it}$ indicate whether the unit $i$ has received the treatment at time $t$. $\mathrm{X}_i$ represent a length $p$ vector of pre-treatment control variables for the unit $i$. Let $T = \{t_0,  \dots, t_M\}$ denote the observed period of time, $\mathrm{Y}_i$ and $\mathrm{Z}_i$ represent the vector of responses and treatment indicators for the unit $i$, respectively. We use lowercase letters, such as $y_{it}$ and vector $\mathrm{y}_i$, to denote the observed values. The data will consist of $n$ independent observations $(\mathrm{y}_i, \mathrm{z}_i, \mathrm{x}_i)$.



In a staggered roll-out design, we consider a unit remains treated once it has received treatment. Within this context, we define $S_{it} = \sum_{s \leq t} Z_{is}$ to represent the time elapsed since treatment adoption, also referred to as the post-treatment period in this study. Treatments assigned at the same time would be a special case in the staggered roll-out design.

The dynamic potential outcomes framework was first introduced by \citep{Robins1986ANA, Robins1987AddendumT}. This framework has been adapted for staggered adoption designs in works such as \citep{Sianesi2004, heckman2016, CALLAWAY2021200}. Under this framework, we use $Y_{it}(S)$ to denote the potential outcome of unit $i$ at time $t$ after being treated for $S$ periods. In the case where $S = 0$, $Y_{it}(0)$ denotes the potential outcome without treatment. The realized treatment effect is then calculated as 
\begin{equation}
Y_{it} = Y_{it}(0) + \sum_{S = 1}^{T}\mathbbm{1}(S_{it} = S) (Y_{it}(S) - Y_{it}(0)).
\end{equation}

For our proposed method, we rely on the SUTVA assumption and extend the strong ignorability assumption to the panel data settings. Namely, when there are no unmeasured confounders, the potential outcomes are independent of the treatment variable conditioning on observed confounders:
\begin{equation}
    \label{eq:ignorability}
    Y_{it}(S) \indep \mathrm{Z}_{i} | \mathrm{X}_i, \quad S = 0, 1, \dots,
\end{equation}
and that every individual has a non-zero probability of adopting the treatment if it is not treated yet:
\begin{equation}
    \label{eq:overlap}
    0 < \text{Pr}(Z_{it} = 1 | \mathrm{X}_i, Z_{i(t-1)} = 0) < 1
\end{equation}

Following the above assumptions, the treatment effect originally defined by the potential outcome framework can be relaxed to the difference in observable outcomes conditioning on the confoundings and time since treatment:
\begin{align*}
    \tau_t(\mathrm{X}_i, S) &:= \mathbb{E}[Y_{it}(S) | \mathrm{X}_i] - \mathbb{E}[Y_{it}(0)|\mathrm{X}_i] \\
    &= \mathbb{E}[Y_{it} | \mathrm{X}_i, S_{it} = S] - \mathbb{E}[Y_{it} | \mathrm{X}_i, S_{it} = 0].
\end{align*}

Expanding from the XBCF model, we assume that the time-varying response to a binary treatment can be approximated by the following equation:
\begin{equation}
    Y_{it} = \alpha \mu(\mathrm{X}_i, T = t) + \beta_S \nu(\mathrm{X}_i,S_{it}, T = t) + \epsilon_{it}.
    \label{eq:longbet}
\end{equation}
The first term captures the prognostic effect in the absence of the treatment. The second term models the dynamic treatment effect as a function of the confoundings, the time since treatment and time $T$. We also introduce a Gaussian process factor $\beta_S$ that captures the general trend of the treatment effect shared among all units and varies by the time since treatment. For simplicity, we assume the error term to be independent Gaussian across units and time. The prognostic effect function $\mu(x_i, t)$ and treatment effect function $\nu(x_i, t, s)$ are modeled by two separate XBART forests, which are modified to handle panel data and consider splits on the time dimensions.

Under the model specification, the treatment effect estimator is given by
\begin{equation}
    \tau_t(\mathrm{X}_i, S) = \beta_S \nu(\mathrm{X}_i, S_{it}, T = t) - \beta_0 \nu(\mathrm{X}_i, S = 0, T = t).
\end{equation}
The treatment effect estimate is obtained by averaging the posterior draws of the estimator, and the method also provides credible intervals for the treatment effect, conditioning on any covariates.

\section{Simulation Study}
To evaluate the performance of our proposed model, we adapt the simulation studies from \citep{richard2020bcf} to generate panel data conditioning on cross-sectional confoundings. We consider a total of $4$ data generating process with the combination of $2$ types of prognostic effects (parallel and non-parallel) and $2$ types of treatment effects (homogeneous and heterogeneous).

We consider $5$ covariate variables which could affect the response variable and treatment adoption but do not change over time. The first three covariate variables are generated from the independent standard normal distribution $x_1, x_2, x_3 \overset{\text{i.i.d}}{\sim} N(0, 1)$; the fourth is a binary variable from binomial distribution; the last one is an unordered categorical variable with three levels and equal probability.
 
The response variable consists of a prognostic effect, a treatment effect (if being treated), and a noise term in the following form:
\begin{equation}
    y(\x_i, t, s) = \eta(\x_i, t) + \mathbbm{1}\{s > 0\}\nu(\x_i, t, s) + \epsilon_{it}, \quad \epsilon_{it}\sim N(0, 0.25)
\end{equation}

When the parallel trend holds, the prognostic effect is the addition of a time factor $f_t$ and a loading factor $\gamma(\x)$ determined by cross-sectional covariates. When the parallel trend is violated, the two factors intertwine to determine the prognostic effect dynamically. 
\[
\eta(\x, t) = \begin{cases}
    f_t + \gamma(\x) \quad & \text{parallel} \\
    f_t \gamma(\x) \quad  & \text{non-parallel.}
\end{cases}
\]
The loading factor is defined as $\gamma(\x) =  g(x_4) + + x_1 | x_3 - 1 |$, where $g(0) = 2$ and $g(1) = -1$. 
The time factor $f_t$ is generated from an ARIMA$(1, 0, 1)$ model with an autoregressive coefficient (ar) of $0.7$ and a moving average coefficient (ma) of $-0.4$ and $1$ is added to each observation to center the time factor around the value 1. 

The treatment effect is composed of a cohort effect, a time factor depending on post-treatment time, and the covariates when considering heterogeneous treatment effect.
\[
\nu(\x, t, s) = \begin{cases}
    2c_{e_i}h_s \quad & \text{homogeneous} \\
    (2 +  x_2 x_5)c_{e_i}h_s \quad & \text{heterogeneous.} 
\end{cases}
\]
The cohort effect $c_{e_i}$ is determined by event time $e_i$ which is when the unit $i$ adopts the treatment. The values of $c_{e_i}$ are listed in Table \ref{tab:cohort_effect} simulating the decaying treatment effect in many product adoption scenarios. The time factor for the treatment effect is defined as $h_s =  se^{-s}$ to mimic the decaying treatment effect after adoption. 

\begin{table}[h!]
\centering
\begin{tabular}{c|cccccc}
\toprule
Event Time $e_i$ & 7 & 8 & 9 & 10 & 11 & 12  \\ \hline
Cohort Effect $c_{e_i}$ & 1.3 & 1.2 & 1.1 & 1.0 & 0.9 & 0.8 \\
\bottomrule
\end{tabular}
\caption{Cohort effect $c_{e_i}$ for each cohort that adopts the treatment at time $e_i = 7, 8, \dots, 12$.}
\label{tab:cohort_effect}
\end{table}

Our simulations generate a total of $T = 12$ time periods with $6$ pre-treatment periods and the treatment starts rolling-out at $t = 7$. Each untreated unit has a chance of adopting the treatment based on their propensity score, which is defined as 
$$\pi(\x) = \frac{1}{5} \Phi(\frac{1}{2}\gamma(x) - \frac{1}{2} x_1)^2 + u_i / 10, $$
where $u_i \sim \text{Uniform}(0, 1)$ and $\Phi(\cdot)$ is the quantile function of standard normal. Starting at $t = 7$, each untreated unit has the probability of $\pi(\x)$ of adopting the treatment.

We assess model performance from three aspects: 1) model accuracy; 2) model reliability: how often the credible interval covers the truth; and 3) decision impact: how well the model can inform decision making. Model accuracy is evaluated through the average root mean squared error (RMSE) calculated across ATT for each cohort and time point. We then assess the average coverage of the 95\% simultaneous interval on the ATTs. In terms of decision impact, we examine how frequently the confidence or credible intervals encompass zero for a non-zero treatment effect, which serves as an indicator of the method's capacity to provide decision-relevant insights. For the LongBet model, we can further measure the RMSE and coverage results specifically for the conditional average treatment effect on the treated (CATT).\footnote{Under the strong ignorability assumption, Longbet can estimate the average treatment effect. However, for consistency with existing literature, we report only the estimated treatment effects on the treated.}

Let $ATT_{e, t}$ represent the average treatment effect on the treated (ATT) for cohort $e$ at time $t$. This results in a total of $21$ ATT estimands across $6$ cohorts and $6$ post-treatment periods. Denote the estimated ATT for cohort $e$ at time $t$ as $\widehat{ATT}_{e, t}$. Furthermore, let $q^{-}_{C, \alpha/2}\{\widehat{ATT}_{e, t}\}$ and $q^{+}_{C, \alpha/2}\{\widehat{ATT}_{e, t}\}$ represents the lower and upper bounds of the 95\% simultaneous confidence interval (or credible interval for Bayesian models) for the $C = 21$ ATT estimands across cohorts and times. With 21 estimands in our scenario, the 95\% simultaneous interval equates to a 99.76\% interval for each estimand, as adjusted by Sidak. The evaluation metrics for ATTs are formulated as follows. We report the metrics averaging over 100 Monte Carlo simulations, offering a thorough analysis of the method's efficacy.

\begin{align*}
\text{RMSE} &= \sqrt{\frac{1}{C} \sum_{e = t_0 + 1}^{t_1} \sum_{t = e}^{t_1} (ATT_{e,t} - \widehat{ATT}_{e, t} )^2} \\
\text{Coverage} &= \frac{1}{C} \sum_{e = t_0 + 1}^{t_1} \sum_{t = e}^{t_1} \mathbbm{1} \left(q^{-}_{C, \alpha/2}\{\widehat{ATT}_{e, t}\} \leq ATT_{e,t} \leq q^{+}_{C, \alpha/2} \{\widehat{ATT}_{e, t}\} \right) \\
\text{Cover0} &= \frac{1}{C} \sum_{e = t_0 + 1}^{t_1} \sum_{t = e}^{t_1} \mathbbm{1} \left(q^{-}_{C, \alpha/2}\{\widehat{ATT}_{e, t}\} \leq 0 \leq q^{+}_{C, \alpha/2} \{\widehat{ATT}_{e, t}\} \right) \\
C &= \frac{2}{(t_1 - t_0 + 1)(t_1 - t_0)} \\
\end{align*}

For LongBet, we set the number of iterations to $120$ ({\tt num\_sweeps}) and ${\tt num\_burnin} = 20$. In each iteration, we train $50$ prognostic trees ({\tt num\_trees\_pr}) and $50$ treatment trees ({\tt num\_trees\_trt}). We compare LongBet to the following baseline methods:
\begin{itemize}
    \item{\textbf{DiD}} This is the difference-in-differences method proposed by \citep{CALLAWAY2021200}. It is tailored for staggered adoption design and relaxes the parallel trend assumption to hold only after conditioning on observed covariates. We employ their method's doubly-robust estimand in our simulations.
    \item{\textbf{Non-linear DiD}} Considering the simulated prognostic function exhibits non-linearity, we incorporate the known relationship $x_1 x_3$ into the DiD estimator to compare its performance with the original version.
    \item{\textbf{DiD Imputation}} Introduced by \citep{borusyak2023revisiting}, the DiD Imputation method involves imputing potential outcomes using observed untreated responses within a fixed effect model framework. This technique is capable of estimating a weighted Average Treatment Effect on the Treated in scenarios featuring heterogeneous causal effects, assuming parallel trends and the absence of anticipation effects. \footnote{The {\tt did\_imputation} package in {\tt R} does not support fitting with continuous cross-sectional covariates. Therefore, we proceeded to fit the model solely with the inclusion of unit- and time-fixed effects.}. 
\end{itemize}

\begin{table}[h!]
\centering
\begin{tabular}{c|rrr|rrr}
\toprule
		\multirow{2}{*}{Method} & \multicolumn{3}{c|}{ATT}  & \multicolumn{3}{c|}{CATT} \\		         & RMSE  & Coverage & Cover 0  & RMSE  & Coverage & Cover 0        \\
\toprule
		\multicolumn{7}{c}{Parallel, Homogeneous}    \\ 
\hline
LongBet & 0.022 & 0.944 & 0.078 & 0.031 & 0.963 & 0.149 \\ 
DiD & 0.024 & 1.000 & 0.130 & - & - & - \\ 
Non-linear DiD & 0.024 & 0.999 & 0.132 & - & - & - \\ 
DiD Imputation & 0.017 & 0.999 & 0.095 & - & - & - \\ 
\toprule
		\multicolumn{7}{c}{Parallel, Heterogeneous}    \\ 
\hline
LongBet & 0.019 & 0.965 & 0.074 & 0.068 & 0.979 & 0.308 \\ 
DiD & 0.023 & 1.000 & 0.140 & - & - & - \\ 
Non-linear DiD & 0.023 & 1.000 & 0.140 & - & - & - \\ 
DiD Imputation & 0.017 & 0.999 & 0.099 & - & - & - \\ 
\toprule
		\multicolumn{7}{c}{Non-parallel, Homogeneous}    \\ 
\hline
LongBet & 0.027 & 0.953 & 0.114 & 0.055 & 0.965 & 0.202 \\ 
DiD & 0.314 & 0.616 & 0.228 & - & - & - \\ 
Non-linear DiD & 0.105 & 0.977 & 0.255 & - & - & - \\
DiD Imputation & 1.337 & 0.041 & 0.138 & - & - & - \\ 
\toprule
		\multicolumn{7}{c}{Non-parallel, Heterogeneous}    \\ 
\hline
LongBet & 0.024 & 0.975 & 0.105 & 0.093 & 0.982 & 0.361 \\ 
DiD & 0.296 & 0.690 & 0.239 & - & - & - \\ 
Non-linear DiD & 0.098 & 0.990 & 0.265 & - & - & - \\ 
DiD Imputation & 1.322 & 0.100 & 0.149 & - & - & - \\ 
\bottomrule
\end{tabular}

\caption{Simulation results}
\label{tab:simulation}
\end{table}

The simulation results, as summarized in Table \ref{tab:simulation}, showcase the performance of the examined methods in simulated scenarios. LongBet stands out as the frontrunner in terms of RMSE on ATT. Remarkably, it achieves competitive coverage on ATT while maintaining a low probability of covering 0. This implies that LongBet is more likely to provide valuable insights for decision-making purposes. Additionally, LongBet boasts low RMSE and high coverage on CATT, reinforcing its proficiency. It's worth noting that LongBet occasionally presents lower coverage while achieving smaller RMSE, owing to its tighter credible intervals that sometimes narrowly miss the true ATT.

In scenarios adhering to the parallel trend assumption, baseline methods demonstrate competitive ATT estimation and higher coverage of the true treatment effect. However, they also have higher coverage on $0$, indicating lower statistical power to reject the ``null hypothesis'' - there is not enought evidence to support the treatment effect is not zero.

On the other hand, when the assumption of parallel trends is not met, methods such as DiD and DiD Imputation, which rely on this assumption, fail to accurately estimate treatment effect. In contrast, the adapted version of DiD demonstrates performance comparable to LongBet. However, there are certain considerations to keep in mind. The non-linear DiD method requires prior knowledge of the true relationship between the response variable and the covariates, which is rarely available in real-world scenarios.

The simulation findings consistently emphasize LongBet's effectiveness in accurately estimating treatment effects and establishing credible intervals across diverse scenarios. Moreover, these simulations bring to light LongBet's strength in situations where the prevalent parallel trend assumption, which forms the basis of numerous existing methods for causal panel data, is not applicable.

\subsection{Prospective Assessment of Treatment Effects with LongBet}
\label{sec:longbet_prospective_assessment}

In the realm of causal inference, a comprehensive understanding extends beyond historical treatment effects to include prospective analyses. A model's capability to anticipate future treatment effects can significantly influence strategic planning and policy development. LongBet, our innovative method, is designed to bridge this analytical divide by facilitating the forward projection of treatment effects.

The LongBet framework encompasses various methodologies to project treatment effects into forthcoming periods. The primary technique involves extrapolating the treatment effect parameter $\beta$, employing Gaussian process regression to discern temporal trends and patterns. Alternatively, LongBet can be amalgamated with the local Gaussian process method detailed in \citep{mwang2023}, which utilizes treatment trees to enhance prediction accuracy. The current discourse focuses on the primary extrapolation technique, evaluating its predictive validity. Investigations into the latter method are earmarked for subsequent research endeavors.

\begin{figure}[h!]
    \centering
    \includegraphics[width=\textwidth]{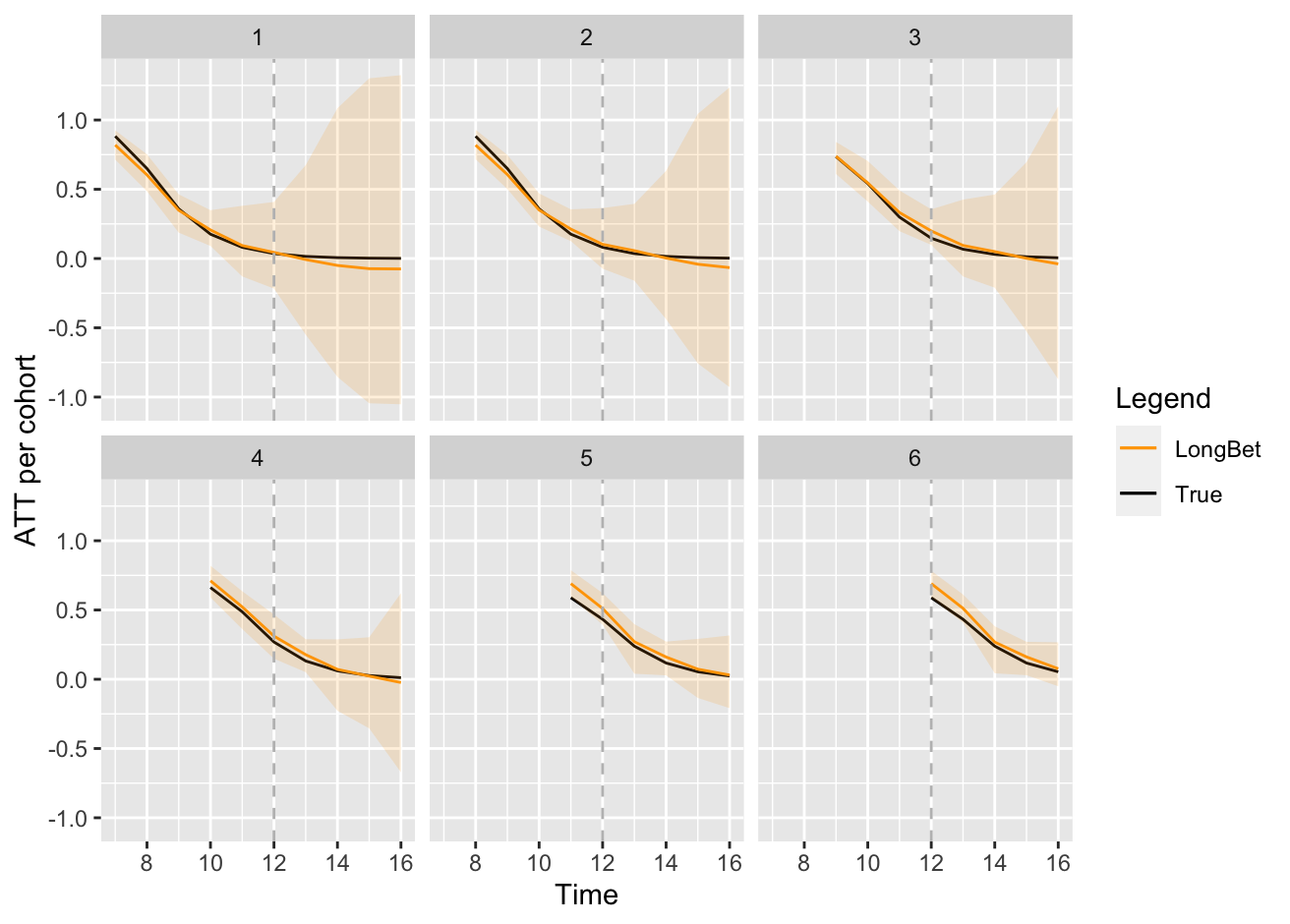}
    \caption[LongBet's Average Treatment Effect Estimation and Forecasting]{LongBet's Average Treatment Effect Estimation and Forecasting. This illustration delineates LongBet's performance up to the 12th period (observed data) and its subsequent forecasts for four additional periods. Each subplot corresponds to a cohort characterized by its initial treatment period. The black trajectory signifies the ground truth of treatment effects, while the grey vertical dashed line demarcates the shift from historical data to LongBet's forecasts. The orange trajectory and the surrounding ribbon depict LongBet's treatment effect estimates and the accompanying $95\%$ credible interval for each respective cohort.}
    \label{fig:longbet_extra_demo}
\end{figure}

This study augments the simulation setup previously introduced, extending the observation window by four periods for future effect estimation. Each cohort's response to treatment is anticipated to exhibit a staggered effect, a phenomenon deftly estimated by LongBet.

As visualized in Figure \ref{fig:longbet_extra_demo}, the delineation between historical and forecasted data is accentuated by a grey vertical line, providing a clear transition from empirical data to LongBet's extrapolations.  The black line corresponds to the ground truth. It is observed that during the observed periods, LongBet's predictions closely align with the ground truth, demonstrating its effectiveness in delineating cohort-specific treatment trajectories.

In the predictive phase, LongBet's estimates display an expansion of the credible intervals, reflecting increased predictive uncertainty. However, for cohorts treated later in the sequence, the intervals are notably narrower, benefiting from the inference drawn from previous cohorts' observations. We have set the default parameters $\sigma = 1$ and $\lambda = (t_1 - t_0)/2$ for the Gaussian process extrapolation. These parameters can be fine-tuned, potentially with domain expertise, to calibrate the prediction uncertainty.  Despite this, LongBet maintains estimations within a pragmatic range.

\section{Conclusion}
In this paper, we have introduced a novel approach for estimating the average and conditional average treatment effects in the context of panel data. Unlike conventional approaches in economics that heavily rely on the parallel trend assumption, our proposed model does not necessitate such an assumption. Instead, it leverages observed covariates to impute potential outcomes and identify treatment effects. The method we presented is grounded in a Bayesian semi-parametric framework, building upon the Bayesian causal forest model and extended to suit the dynamics of panel data settings. The method combines the strengths of Bayesian methodologies and offers credible intervals for conditional average treatment effects. 

It is important to acknowledge the underlying assumption of no unobserved confounding in LongBet, which may not be universally applicable. However, the proliferation of large-scale datasets in the era of big data offers potential solutions to address this limitation. The semi-parametric nature of our approach empowers it to capture complex and non-linear relationships between covariates and responses, which can lead to more accurate treatment effect estimation than traditional methods.

Looking ahead, our approach offers promising avenues for future research. Extending LongBet to encompass time-varying covariates is a logical step forward, allowing for a more nuanced understanding of how treatments interact with changing covariate profiles over time. Additionally, extrapolating the estimated causal effects by LongBet to provide predictive inferences for the future could yield valuable insights for decision-makers.

In conclusion, LongBet emerges as a promising tool for causal inference in panel data settings, offering a departure from conventional assumptions while providing credible and insightful treatment effect estimates. As we acknowledge its potential limitations and embrace its strengths, we envision a dynamic landscape of research and application, where LongBet continually evolves to address complex real-world scenarios and contributes to the advancement of causal inference in panel data analysis.

\bibliographystyle{apalike}
\bibliography{bib}

\begin{thebibliography}{}

\bibitem[Abadie, 2005]{Abadie2005SemiDID}
Abadie, A. (2005).
\newblock {Semiparametric Difference-in-Differences Estimators}.
\newblock {\em The Review of Economic Studies}, 72(1):1--19.

\bibitem[Abadie, 2021]{Abadie2021UsingSC}
Abadie, A. (2021).
\newblock Using synthetic controls: Feasibility, data requirements, and
  methodological aspects.
\newblock {\em Journal of Economic Literature}.

\bibitem[Abadie and L’Hour, 2021]{abadie2021penalizedsc}
Abadie, A. and L’Hour, J. (2021).
\newblock A penalized synthetic control estimator for disaggregated data.
\newblock {\em Journal of the American Statistical Association},
  116(536):1817--1834.

\bibitem[Acemoglu et~al., 2013]{Acemoglu2013TheVO}
Acemoglu, D., Johnson, S., and Kermani, A. P.~P. (2013).
\newblock The value of connections in turbulent times: Evidence from the united
  states.
\newblock {\em Financial Crises eJournal}.

\bibitem[Athey et~al., 2021]{Athey_2021}
Athey, S., Bayati, M., Doudchenko, N., Imbens, G., and Khosravi, K. (2021).
\newblock Matrix completion methods for causal panel data models.
\newblock {\em Journal of the American Statistical Association},
  116(536):1716--1730.

\bibitem[Bernal et~al., 2016]{Bernal2016InterruptedTS}
Bernal, J.~L., Cummins, S., and Gasparrini, A. (2016).
\newblock Interrupted time series regression for the evaluation of public
  health interventions: a tutorial.
\newblock {\em International Journal of Epidemiology}, 46:348 -- 355.

\bibitem[Bernal et~al., 2018]{LopezBernal2018AMF}
Bernal, J.~L., Soumerai, S.~B., and Gasparrini, A. (2018).
\newblock A methodological framework for model selection in interrupted time
  series studies.
\newblock {\em Journal of clinical epidemiology}, 103:82--91.

\bibitem[Borusyak et~al., 2023]{borusyak2023revisiting}
Borusyak, K., Jaravel, X., and Spiess, J. (2023).
\newblock Revisiting event study designs: Robust and efficient estimation.

\bibitem[Callaway and Karami, 2023]{CALLAWAY2023184}
Callaway, B. and Karami, S. (2023).
\newblock Treatment effects in interactive fixed effects models with a small
  number of time periods.
\newblock {\em Journal of Econometrics}, 233(1):184--208.

\bibitem[Callaway and Sant’Anna, 2021]{CALLAWAY2021200}
Callaway, B. and Sant’Anna, P.~H. (2021).
\newblock Difference-in-differences with multiple time periods.
\newblock {\em Journal of Econometrics}, 225(2):200--230.
\newblock Themed Issue: Treatment Effect 1.

\bibitem[Chipman et~al., 2010]{chipman2010bart}
Chipman, H.~A., George, E.~I., and McCulloch, R.~E. (2010).
\newblock {BART: Bayesian additive regression trees}.
\newblock {\em The Annals of Applied Statistics}, 4(1):266 -- 298.

\bibitem[Dube and Zipperer, 2015]{dube2015poolsc}
Dube, A. and Zipperer, B. (2015).
\newblock Pooling multiple case studies using synthetic controls: An
  application to minimum wage policies.
\newblock {\em IZA Discussion Paper}, (8944).

\bibitem[Hahn et~al., 2020]{richard2020bcf}
Hahn, P.~R., Murray, J.~S., and Carvalho, C.~M. (2020).
\newblock {Bayesian Regression Tree Models for Causal Inference:
  Regularization, Confounding, and Heterogeneous Effects (with Discussion)}.
\newblock {\em Bayesian Analysis}, 15(3):965 -- 2020.

\bibitem[He and Hahn, 2023]{he2023stochastic}
He, J. and Hahn, P.~R. (2023).
\newblock Stochastic tree ensembles for regularized nonlinear regression.
\newblock {\em Journal of the American Statistical Association},
  118(541):551--570.

\bibitem[He et~al., 2019]{he2019xbart}
He, J., Yalov, S., and Hahn, P.~R. (2019).
\newblock Xbart: Accelerated bayesian additive regression trees.
\newblock In Chaudhuri, K. and Sugiyama, M., editors, {\em Proceedings of the
  Twenty-Second International Conference on Artificial Intelligence and
  Statistics}, volume~89 of {\em Proceedings of Machine Learning Research},
  pages 1130--1138. PMLR.

\bibitem[Heckman et~al., 2016]{heckman2016}
Heckman, J., Humphries, J., and Veramendi, G. (2016).
\newblock Dynamic treatment effects.
\newblock {\em Journal of Econometrics}, 191(2):276--292.

\bibitem[Hill, 2011]{hill2011bayesian}
Hill, J.~L. (2011).
\newblock Bayesian nonparametric modeling for causal inference.
\newblock {\em Journal of Computational and Graphical Statistics},
  20(1):217--240.

\bibitem[Imbens et~al., 2021]{imbens2021controlling}
Imbens, G., Kallus, N., and Mao, X. (2021).
\newblock Controlling for unmeasured confounding in panel data using minimal
  bridge functions: From two-way fixed effects to factor models.

\bibitem[Imbens and Rubin, 2015]{Imbens2015CausalIF}
Imbens, G. and Rubin, D.~B. (2015).
\newblock Causal inference for statistics, social, and biomedical sciences:
  Sensitivity analysis and bounds.

\bibitem[Krantsevich et~al., 2022]{Krantsevich2022StochasticTE}
Krantsevich, N.~M., He, J., and Hahn, P.~R. (2022).
\newblock Stochastic tree ensembles for estimating heterogeneous effects.
\newblock In {\em International Conference on Artificial Intelligence and
  Statistics}.

\bibitem[Kreif et~al., 2016]{kreif2016sc}
Kreif, N., Grieve, R., Hangartner, D., Turner, A.~J., Nikolova, S., and Sutton,
  M. (2016).
\newblock Examination of the synthetic control method for evaluating health
  policies with multiple treated units.
\newblock {\em Health Economics}, 25(12):1514--1528.

\bibitem[Robins, 1986]{Robins1986ANA}
Robins, J.~M. (1986).
\newblock A new approach to causal inference in mortality studies with a
  sustained exposure period—application to control of the healthy worker
  survivor effect.
\newblock {\em Mathematical Modelling}, 7:1393--1512.

\bibitem[Robins, 1987]{Robins1987AddendumT}
Robins, J.~M. (1987).
\newblock Addendum to “a new approach to causal inference in mortality
  studies with a sustained exposure period—application to control of the
  healthy worker survivor effect”.
\newblock {\em Computers \& Mathematics With Applications}, 14:923--945.

\bibitem[Schaffer et~al., 2021]{Schaffer2021InterruptedTS}
Schaffer, A.~L., Dobbins, T.~A., and Pearson, S.-A. (2021).
\newblock Interrupted time series analysis using autoregressive integrated
  moving average (arima) models: a guide for evaluating large-scale health
  interventions.
\newblock {\em BMC Medical Research Methodology}, 21.

\bibitem[Sianesi, 2004]{Sianesi2004}
Sianesi, B. (2004).
\newblock An evaluation of the swedish system of active labor market programs
  in the 1990s.
\newblock {\em The Review of Economics and Statistics}, 86(1):133--155.

\bibitem[Tanaka, 2020]{tanaka2020bayesian}
Tanaka, M. (2020).
\newblock Bayesian matrix completion approach to causal inference with panel
  data.

\bibitem[Wang et~al., 2023]{mwang2023}
Wang, M., He, J., and Hahn, P.~R. (2023).
\newblock Local gaussian process extrapolation for bart models with
  applications to causal inference.
\newblock {\em Journal of Computational and Graphical Statistics}, 0(0):1--12.

\bibitem[Wooldridge, 2021]{jeffrey2021twmr}
Wooldridge, J. (2021).
\newblock Two-way fixed effects, the two-way mundlak regression, and
  difference-in-differences estimators.
\newblock {\em SSRN Electronic Journal}.

\bibitem[Xu, 2017]{xu_2017}
Xu, Y. (2017).
\newblock Generalized synthetic control method: Causal inference with
  interactive fixed effects models.
\newblock {\em Political Analysis}, 25(1):57–76.

\end{thebibliography}

\end{document}